# Dynamically tunable electromagnetically induced transparency-like metamaterial structure based on polarization sensitivity

KE DI,[1,2,3] MENG XIE,[1,2,4] ZHAOYANG WANG,[1,2,5] RENPU LI,[1,2,6] YU LIU,[1,2,7] AND JIAJIA DU[1,2,*]

[1]*Department of Optoelectronic Engineering, Chongqing University of Post and Telecommunications, Chongqing 400065, China*

[2]*Chongqing Key Laboratory of Autonomous Navigation and Microsystem, Chongqing University of Posts and Telecommunications,Chongqing 400065, China*

[3]dike@cqupt.edu.cn

[4]s210431109@stu.cqupt.edu.cn

[5]1416809852@qq.com

[6]lirp@cqupt.edu.cn

[7]liuyu@cqupt.edu.cn

[*]dujj@cqupt.edu.cn

**Abstract:** In this paper, we propose a plasmon-induced transparency (PIT) metamaterial structure composed of Ag nanomaterials with polarization sensitivity. The metamaterial model consists of three bright modes with different resonant frequencies. The optical properties of the structure are further investigated using finite difference time domain (FDTD) method. The results show that the conversion between single-band PIT and dual-band PIT effects can be achieved by changing the polarization degree of the incident light, the number of transparent windows can be changed from one to two, and the process is accompanied by the conversion of bright and dark modes and the change of the resonance wavelength of the transmission peak. In addition, When the light is polarized in the Y-direction, the two transparency windows have different refractive index sensitivities, with *FOM* values of 5.94/RIU and 5.65/RIU, respectively.

## 1. Introduction

Electromagnetically induced transparency (EIT) is a quantum interference effect in three-level atomic system. The interaction of the effect is through the interference of two different quantum transition channels, producing a "transmission window" with a certain width in the broad absorption spectrum. In 1989, the Harris team proposed an early theory of EIT, and the team first observed the EIT phenomenon in 1990[1]. However, the EIT phenomenon requires harsh experimental conditions such as low temperature and high-intensity pumped light source in atomic system, which limits its deeper development and applications. In 2008, Zhang et al [2]. designed a periodic metamaterial structure consisting of three metal bars. The team observed for the first time an EIT-like effect at room temperature based on this material, which is free from harsh experimental conditions. This kind of EIT-like effect is known as Plasmon Induced Transparency (PIT). EIT effect based on metamaterials can be observed in the visible [3,4] , near-infrared [5,6], terahertz [7-9] and radio frequency bands [10-12]. Therefore, the research of metamaterial-based EIT effect has a wide range of prospects for applications in micronano-photonic devices such as high-sensitivity optical sensor[13],slow-light caching[14], and narrow band filter[15].

Bright-dark mode coupling and bright-bright mode coupling are two implementation methods of the EIT effect in metamaterials [16]. Bright mode is a mode in which a structure is excited directly by the incident field and then in a resonant state. Dark mode entry into excited state requires near-field action of bright mode [17]. A metamaterial structure consisting of a graphene disk and a graphene strip was designed to obtain the single-band PIT effect in X-polarized light by bright-dark mode coupling [18]. A metamaterial structure consisting of cross-shaped and 4L-shaped graphene patterns also realizes a single-band PIT effect through bright-bright mode coupling [19]. However, multi-band PIT has the advantage of higher flexibility and accuracy compared to single-band PIT because it carries more frequencies. Therefore, the multi-band PIT phenomenon based on metal resonant structures has becoming a key research topic nowadays [20-22]. A metamaterial structure designed using gold material with high refractive index sensitivity for PIT effect in dual-band is proposed [23]. The two-band PIT effect in X-polarized light was obtained using two pairs of split-ring resonators and a cut wire, and the slow-light effect and the energy storage capacity of the window were experimentally verified [24]. However， most metamaterials focus on only one polarization direction of the source, and less attention is paid to controlling the EIT effect using the polarization direction of the source.

In this paper, we propose a dynamically tunable EIT-like metamaterial model. We realize the conversion between single-band PIT and dual-band PIT effect by varying the polarization angle of the incident light. The spectra and electric field maps of the structures are obtained through software simulation to further investigate and analyze the PIT generation principle and the influence factors. It was shown that the resonant wavelength and amplitude of the transparency window can be tuned by varying the opening length of the Ag split ring, the length of the Ag cut wire, and the length of the Ag square ring. Finally, we investigate the refractive index sensing performance of transparent windows and demonstrate the promising potential of this design for refractive index sensor applications.

## 2. Structure model and simulation results

The model diagram and geometrical parameters of the structure proposed in this paper are shown in Fig.1. From left to right, the structure consists of an Ag Split ring resonator (SRR), an Ag cut wire (CW), and an Ag square ring (SR), respectively. The width of the SRR opening is defined as a and the length is defined as b. We use a*b to indicate size of the opening. SRR and SR are placed on each side of the CW. The purpose of using SRR is to break the symmetry of the structure and thus bring in a new mode to obtain significant electromagnetically induced transparency phenomenon [25,26]. The specific parameters of the structure are shown in Fig.1(a). The width and length of the CW are defined as $W_1 = 50$ nm and $L_1 = 300$ nm. The length and width of the SR are defined as $L_2 = 150$ nm and $L_3 = 70$ nm. The SRR has a length of $L_4 = 150$ nm and a width of $L_5 = 70$ nm. The distance of the two side structures from the center CW is defined as $g$, $g=25$nm. The size of the substrate material is $P_X = P_Y = 350$nm, and its thickness is $h = 60$ nm. The simulated light source is set to 600 nm-1600 nm, and the direction of the electric field is $\theta$ angle to the X-axis of the metamaterial. The full metamaterial structures are placed as in Fig.1(b) on a substrate with a refractive index of 1.5 and its period set to $P$. Ag is a very effective plasma material, and its parameters are defined by the metal Drude model [27]:

$$\varepsilon_{r(\omega)} = \varepsilon_\infty - \frac{\omega_p^2}{\omega^2 + i\gamma\omega}, \qquad (1)$$

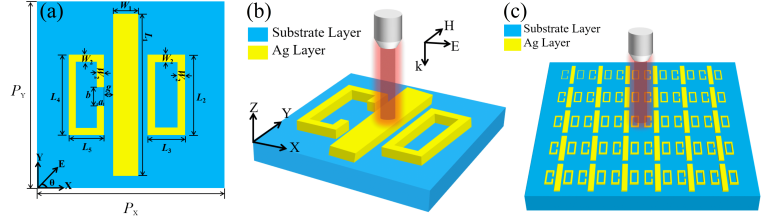

**Fig.1.** Structural model diagram. (a) Two-dimensional metamaterial structural unit. (b) Three-dimensional structural unit. (c) Three-dimensional spatial structure.

Here, $\varepsilon_\infty$ is the relative permittivity of the Ag material at infinite frequency, $\omega_p$ is the plasma resonance frequency, $\gamma$ is the damping factor of the metal, and $\omega$ is the frequency of the incident electromagnetic wave. In this design, the parameters are respectively set as $\varepsilon_\infty = 3.7$, $\omega_p = 1.37 \times 10^{16}$ rad/s, $\gamma = 3.07 \times 10^{13}$ rad/s 。

The design was simulated and validated in the commercial software FDTD Solutions. Since, the model is periodic in the Y and Y directions as in Fig.1(c), we set the boundary conditions in the X and Y directions as periodic boundary and in the Z direction as perfect absorption layer in the simulation.

When $\theta=0°$ (X polarization), the simulated transmission spectrum result is shown in Fig.2(a). In Fig.2(a), SRR and SR can act individually with the incident light into a resonant state, and eventually produce an obvious transmission valley at 797 nm and 852.5 nm, respectively. However, CW has no transmission dip in its transmission spectrum. The transmission spectrum of the whole structure has a transparent window with a wide range of widths at 840nm, up to 84% of the transmission. Transmission valleys and transmission peaks are labeled as A (788.7 nm), B (840 nm), and C (889 nm) in Fig.2(a). Fig.2(b)-(d) demonstrate the electric field distributions of the transmission valleys A and C and transmission peak B. As can be seen from Fig.2(b) and Fig.2(d), SRR and SR accumulate a lot of positive and negative charges at the wavelengths of the two transmission valleys exhibiting dipole modes in a strong resonance state. However, in Fig.2(c), the destructive interference between the structures suppressing the resonance of the SRR is the reason for the formation of the transparent window at B, and the weaker electric field strength here compared to that at the transmission valley is also able to reflect this. This destructive interference suppresses the absorption of the light source by the structure increasing the transmittance and creating a transparent window at 840 nm. The CW is barely involved in the formation of the transparent window. In summary, in the presence of an X-polarized source, both SRR and SR structures act as bright modes, and the interplay of SRR and SR is responsible for the single-band PIT effect.

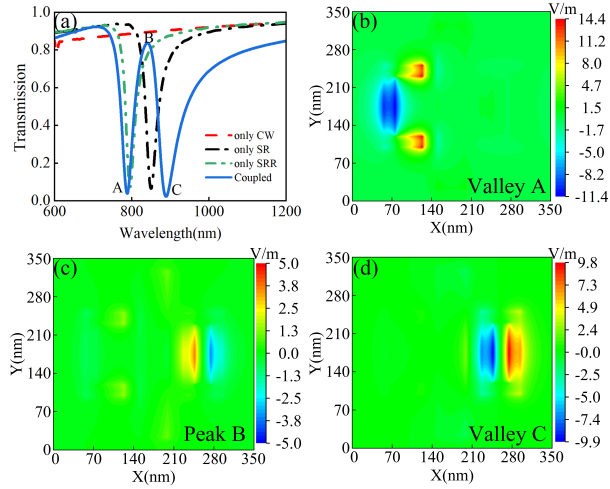

**Fig.2.** Simulation results when the light polarization angle θ=0°. (a) Transmission spectra of SRR, SR, CW and whole structure. The electric field of (b) valley A, (c) peak B, (d) valley C.

## 3. Physical performance analysis

In the next, since the structure is asymmetric, we change the angle of polarization of the incident light, θ, to study its influence on the PIT effect. The results of the transmission spectrum when θ varying from 0° to 90° are shown in Fig.3(a). When the polarization angle θ increases to 90° (Y-polarization), the transparent window at B disappears, meanwhile two transparent windows with different amplitudes are generated at 1000 nm and 1346 nm in Fig.3(b). The new two transparent windows are labeled as E and G in the figure reaching 86% and 54% transmittance, and the three transmission valleys are labeled as D (820.6 nm), F (1156.6 nm), and H (1408 nm).

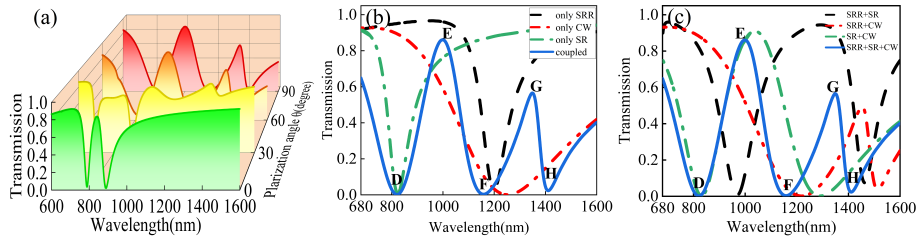

**Fig.3.** Influence of light source polarization angle on PIT effect. (a) The angle of polarization of the light source affects the transmission spectrum. Simulation results of (b) SRR, SR, CW individually and whole structure, (c) CW coupled SR, SRR coupled SR, SRR coupled CW and whole structure when the angle of polarization is θ=90°.

In Fig.3(b), the transmission spectra of SRR, SR and CW are shown simultaneously when they work individually at θ=90°. All three structures are all able to couple to the incident light, which is different from the results in X-polarized light, but their coupling strengths differ. Larger Q-factors indicate weaker coupling to the incident light, so modes with normalized SRR and SR are considered as quasi-dark modes and modes with CW are considered as bright modes. However, it is important to note that SRR and SR are different from true dark mode because they can both interact with the incident field. In order to further understand the origin of dual-band PIT, the transmission spectrum simulation results of SRR+SR, SRR+CW, SR+CW, and SRR+CW+SR are shown in Fig.3(c). The results show that three different structures can be combined by two-by-two to obtain three different single-band PIT. The coupling of the SR and CW structures results in the higher amplitude transparent window. Low amplitude transparency window results from the interaction between SRR and CW. Although, the same coupling effect exists for SRR and SR, the effect is suppressed in the overall structure.

We extracted the electric field profiles at each transmission peak and valley to analyze and validate the above results. Fig.4(a)-(e) show the Z-component of the electric field of the structure at different wavelengths at D, E, F, G, and H, respectively. From Fig.4(a), the resonance of CW and SRR is very weak at the transmission valley D. The charge distribution of SR shows an upward positive and downward negative forming a dipole mode, which is in a very strong resonance state, leading to the transmission dip. At the wavelength of the transmission valley F, the three structures without energy exchange are in resonance and the surface electric field strength is strongly manifested. Because the SRR and CW, which are close in frequency when resonating individually as in Fig.3(c), have a same sign of accumulated charge at their close ends, their phases are the same will attract each other and perform as symmetric mode resonance. The accumulated charges at the SR and CW ends are of opposite sign, and the Coulomb repulsion occurs at their close ends, forming an antisymmetric mode resonance. In contrast, the electric field strength on the structure is significantly suppressed at the transparent with E in the Fig.4(b). This is due to the fact that the resonances of SRR, CW and SR are all suppressed in the destructive interference, yielding a transparent window with a higher transmission coefficient. Similarly, at H wavelength, SRR and CW, which are not exchanging energy but are simultaneously in a resonant state with opposite charges at their ends shown as Fig.4(e), form a transmission valley. But, due to the occurrence of destructive interference the resonance of the CW and SR structures is completely suppressed shown as Fig.4(d), forming a transparent window at G. The SRR structure is also somewhat suppressed but not completely is the cause of the lower amplitude of the window at G. Thus, the transparent windows in different wavelength bands are caused by the interference effect between different structural bodies.

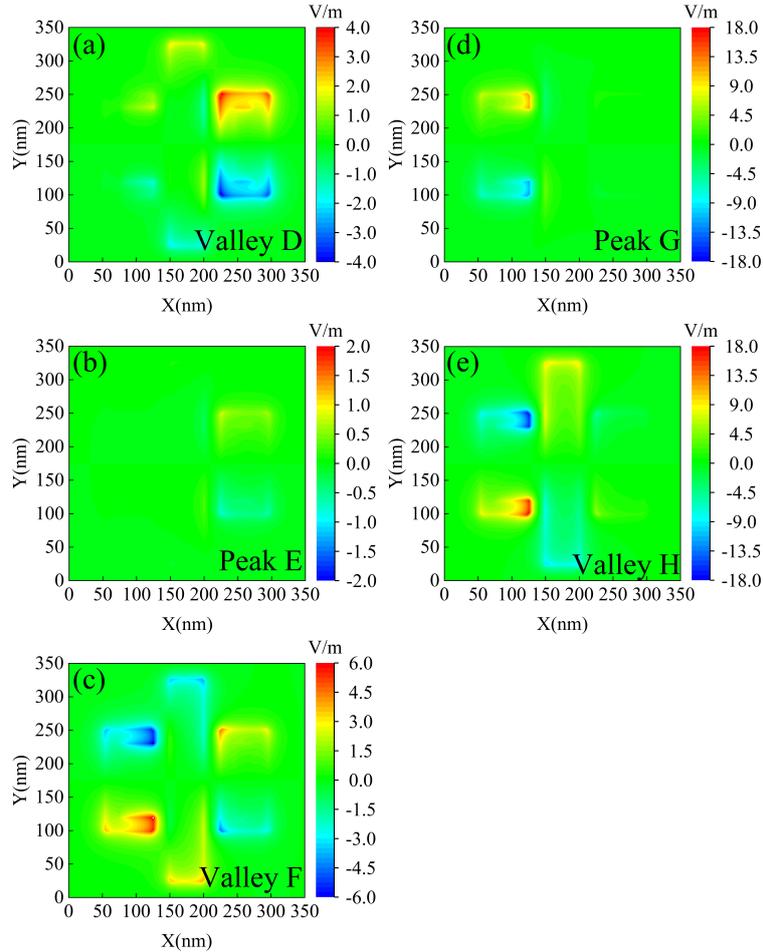

**Fig.4.** Distribution of the Z-component of the electric field when the angle of polarization is $\theta=90°$. The electric field of (a) valley D, (b) peak B, (c) valley F, (d) peak G, (e) valley H.

In summary, when the polarization angle $\theta$ of the light source varies in the range of 0° ~ 90°, the transmission spectrum transforms from a single-band PIT to a dual-band PIT. Moreover, this process involves the transition of the bright and dark modes and the shift of the resonant wavelength of the transmitted peak. This means that the PIT effect can be dynamically tuned in this model by controlling the polarization direction of the light source.

Based on the above analysis, the transparent windows originate from destructive interference of structures. However, the change in structural parameters can lead to a shift in its resonance wavelength and further affect the PIT results. Fig.5 illustrates the three structural parameters that affect the results of the transmission spectra at Y-polarized optical conditions. As shown in Fig.5(a), when the opening size $b$ of the SRR decreases from 110 nm to 20 nm, the transparent window at E still exists, and the transmission valley

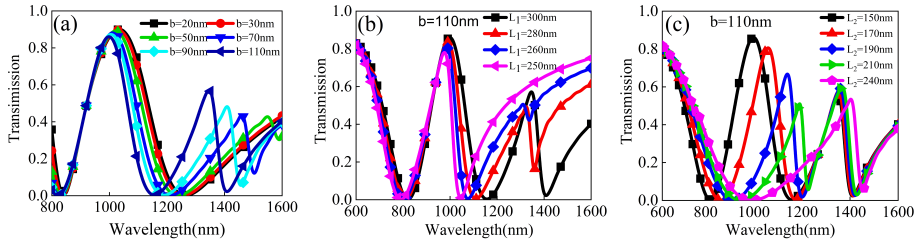

**Fig.5.** The results of the structure parameters affecting the PIT window when the angle of polarization is $\theta$=90°. Results of transmission spectra obtained because of changes in (a) $b$, (b) $L_1$, (c) $L_2$

at F is shifted to further broadens the bandwidth of E. But, the transparent window at G disappears in the process. This is because a smaller opening size of the SRR makes the overall structure more symmetrical. When the opening size is smaller than 20 nm, the resonance frequencies of SRR and SR are so close to each other that they act together as a single mode. Fig.5(b) demonstrates how $L_1$ affects the transmission spectrum. As $L_1$ is decreasing, the transmission valley at F is blue-shifted, and at the same time it narrows the linewidth of the transparent window E. Transmission valley at H gets blue-shifted and amplitude-increased due to increasing $L_1$, which eventually leads to the disappearance of the G transparent window at $L_1 = 250$ nm. Therefore, the opening length b of the SRR and the length $L_1$ of the CW can be adjusted to adjust the transparency window at G, in agreement with the results obtained when analyzing the electric field. Fig.5(c) demonstrates how $L_2$ affects the transmission spectrum. When $L_2$ varies in the range of 150 nm-240 nm, the two transmission valleys D and F are significantly red-shifted at the same time, causing a significant decrease in the amplitude and window width of the transparent window at E. When $L_2$ increases to 240 nm, the transparent window at E gradually disappears, which is due to the longer $L_2$ the longer the dipole distance of the SR, and the resonance frequency of the SR shifts to lower frequency. Hence, tuning $L_2$ can control the transparency window E.

From the above analysis, we can see that changes in the structural parameters can adjust the number of transparent windows in the transmission spectrum.

## 4. Refractive index sensing performance

Metamaterials have great potential for refractive index sensing applications due to their ability to greatly enhance the electric field strength in a small area. Metamaterial resonance in the high frequency band can be equal to plasma resonance [28], and its resonant frequency is defined as:

$$\omega_q \propto \frac{1}{2d\sqrt{\varepsilon_{eff}}} = \frac{1}{2d\sqrt{\Omega\varepsilon_{air} + (1-\Omega)\varepsilon_m + \varepsilon_{sub}}}, \qquad (2)$$

In the above equation, $d$ is the size of the unit structure, $\varepsilon_{eff}$ denotes the equivalent dielectric constant of the environment surrounding the metamaterial. $\varepsilon_{air}$、$\varepsilon_m$ and $\varepsilon_{sub}$ represent the dielectric constant of air, object to be measured on metamaterial surface, and the substrate material, respectively. The percentage of air in the environment is defined as $\Omega$. From the above equation, the resonant frequency of the metamaterial is inversely proportional to the arithmetic square root of the equivalent dielectric constant of its surroundings. We evaluate the performance of the sensor using two important parameters, Sensitivity (*S*) and Figure Of Merit (*FOM*) [29], and they are defined respectively as:

$$S = \frac{\Delta\lambda}{\Delta n}, \qquad (3)$$

$$FOM = \frac{S}{FWHM}, \qquad (4)$$

$\Delta\lambda$ and *FWHM* are the wavelength shift and the half-width of the transmission spectrum, respectively. *S* indicates the amount of wavelength shift due to a unit change in refractive index. The higher sensitivity indicates a greater refractive index detection capability of the structure. *FOM* is a parameter that reflects the overall performance of a sensor.

Fig. 6 demonstrates the results of the refractive index sensing performance of the structure under two polarized lights. Among them, Fig.6(a) shows the change of transmission spectrum due to the change in environmental refractive index under X-polarized light condition. The result shows that when the environmental refractive index increases from 1 to 1.1, the resonant frequency of the metamaterial decreases and the overall transmission spectrum is red-shifted, which is consistent with Eq. (2). The refractive index sensitivity of window B is defined as $S_B$, Fig.6(b) is linear fitting results of environmental refractive index with the wavelength of the transmission peak B, and we get $S_B$ = 385 nm/RIU, and the *FOM* value is 5.62/RIU. From Fig. 5, $L_2$ change in the 150 nm-200 nm range does not affect the number of transparent windows, and meanwhile increasing $L_2$ benefits the SR to confine more electric fields to enhance the sensing performance of the structure. The effect of $L_2$ on $S_B$ is shown in Fig.6(c). As $L_2$ increases, the sensitivity and *FOM* values at B show an opposite trend. The *FOM* value at B is maximum at $L_2$=150 nm with a maximum value of 5.62/RIU. $S_B$ reaches its maximum value at $L_2$=200 nm with $S_B$=530 nm/RIU and a *FOM* value of 1.77/RIU.

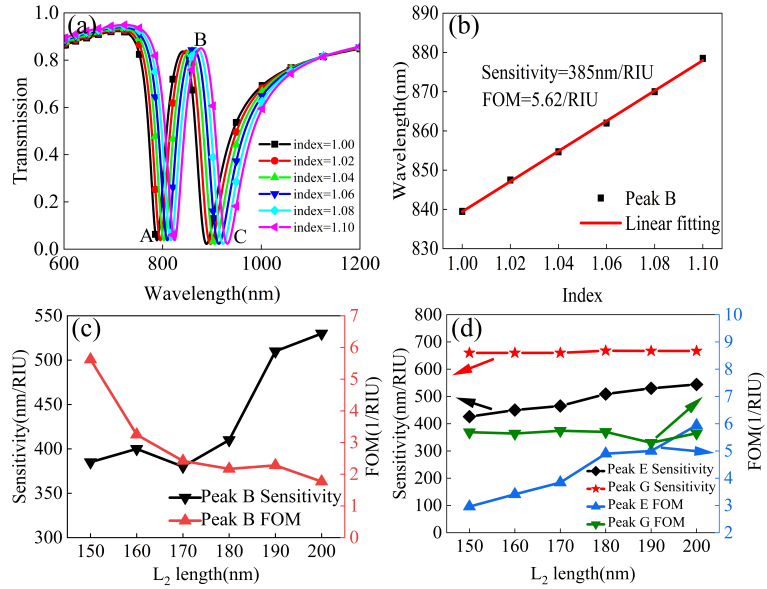

**Fig.6.** Refractive index sensing performance of metamaterials. (a) The relationship between refractive index and transmission spectrum. (b) The results of fitting the resonance wavelength of the transmission peak at B to the ambient refractive index. $L_2$ affects the sensitivity and *FOM* value of (c) peak B. (d) peak E and peak G

The refractive index sensitivities of the two transparent windows in Y-polarized light are defined as $S_E$ and $S_G$, respectively. The relationship of parameter $L_2$ with transmission peak E, G sensitivity and *FOM* values is shown in Fig. 6(d). The increment in $L_2$ leads to an increase in sensitivity and *FOM* values in the transparency window E, while there is no significant change in sensitivity and *FOM* values in the G window. The transparency window of E at $L_2$ = 200 nm has the largest sensitivity and *FOM* values. The maximum $S_E$ value is 544 nm/RIU and the maximum *FOM* value of E is 5.94/RIU. At the same time, the sensitivity of the transmission peak at G reaches 667 nm/RIU with a *FOM* value of 5.65/RIU.

The metamaterial structure design proposed in this paper is not only able to actively adjust the number of transparent windows, which improves the applicability of the design, but also the *FOM* values of the transparent windows can all reach more than 5/RIU. We can improve the detection accuracy of the system by checking the position of the two transparent windows to rule out other disturbances in the actual sensing detection.

## 5. Conclusion

In conclusion, we have presented a metamaterial structure model for Ag materials that exploits the polarization of the light source to achieve a dynamically tunable PIT effect. The Ag material itself is not actively tunable, but the single-band PIT will transform into a dual-band PIT when the incident light is changed from X-polarized to Y-polarized, enabling active tuning of the PIT effect. At the same time, the two transparency windows have different refractive index sensitivities under Y-polarized light, with *FOM* values up to 5.94/RIU and 5.65/RIU respectively. Utilizing a dual-band PIT can more effectively eliminate interference in the actual sensing detection and improve the detection accuracy, thus the metamaterial model provides a reference for refractive index sensors.

**Funding.** Young Scientists Fund of the National Natural Science Foundation of China (Grant No. 11704053), the National Natural Science Foundation of China (Grant No. 52175531) and the Science and Technology Research Program of Chongqing Municipal Education Commission (Grant No. KJQN 201800629，KJZD-M202000602, 62375031).